\documentstyle[seceq,epsf]{ptptex}

\notypesetlogo  

\newcommand{\ac}{ {\alpha_{c}} }
 
\newcommand{\beff}{B_{\rm eff}}
\newcommand{\cm}{ {\rm cm} }

\newcommand{\muu}{ {\mu_{u}} }
\newcommand{\mud}{ {\mu_{d}} } 
\newcommand{\mus}{ {\mu_{s}} }
\newcommand{\mue}{ {\mu_{e}} } 
\newcommand{\order}{{\cal O}}
\newcommand{\pc}{ {\rm pc} }

\newcommand{\simg}
   {\mathrel{\raise.3ex\hbox{$>$\kern-.75em\lower1ex\hbox{$\sim$}}}}
\newcommand{\siml}
   {\mathrel{\raise.3ex\hbox{$<$\kern-.75em\lower1ex\hbox{$\sim$}}}}

\title{
Upper Limit on the Mass of RX J1856.5--3754 \\ as a Possible Quark Star
}

\author{
Kazunori {\sc Kohri},$^{1}$
Kei  {\sc Iida}$^{2}$
and Katsuhiko {\sc Sato}$^{1,3}$
}

\inst{
$^1$Research Center for the Early Universe (RESCEU), School of
Science, \\ University of Tokyo, Tokyo 113-0033, Japan \\ 
$^2$The Institute of Physical and Chemical Research (RIKEN), Wako,
351-0198, Japan \\ 
$^3$Department of Physics, School of Science, University of Tokyo,\\
Tokyo 113-0033, Japan }

\recdate{
October 11, 2002
}

\abst{
  Recent deep Chandra LETG+HRC-S observations suggest the possibility
  that  RX J1856.5--3754 is a compact star whose radiation radius is
  3.8--8.2 km.  In this paper, we systematically calculate the
  mass-radius relations of quark stars within the bag model.  Assuming
  that RX J1856.5--3754 is a pure quark star, we derive an upper
  limit on its mass for various sets of bag-model parameters. 
}

\begin{document}

\maketitle

\section{Introduction}

The possibility that compact stars supported by the degenerate
pressure of quark matter exist has been investigated by many authors
(see, e.g., Refs.\
\citen{ivanenko:1969,itoh:1970,Collins:1974ky,Baym:yu,Chapline:gy,Kislinger:1978,Fechner:ji,Freedman:1977gz,Baluni:1977mk,Witten:1984rs,Alcock:1986hz,Haensel:qb,Rosenhauer,Glendenning,Schertler,Fraga,Andersen,Peshier,Zdunik,Sinha}).
The most commonly accepted picture of such quark stars is that a star
containing quark matter in its core region, surrounded by hadronic
matter, might exist in the branch of neutron stars;\cite{Rosenhauer}
this hypothesized type of star is often referred to as a ``hybrid
star''. If hadronic matter in neutron stars can undergo a strong
first-order phase transition, it is possible that a third family of
more compact stars exists.\cite{Gerlach} Such a third family, arising
from a deconfinement phase transition, is predicted by a very
restricted class of models of quark and hadronic
matter.\cite{Glendenning,Schertler} As suggested by
Witten,\cite{Witten:1984rs} there is another possible form of quark
stars: If the true ground state of hadrons is ``strange matter'', bulk
quark matter consisting of approximately equal numbers of $u$, $d$,
and $s$ quarks, then self-bound quark stars (or ``strange stars'')
could occur at masses and radii on the order of or even smaller than
those of typical neutron star, $\sim10$ km and $\sim1.4M_{\odot}$. In
the absence of reliable information about the equilibrium properties
of hadronic and quark matter at high densities, however, it is
impossible to tell which kind of quark stars is favored.

Recently, Drake et al.\cite{Drake:2002bj} reported that the deep
Chandra LETG+HRC-S observations of the soft X-ray source RX
J1856.5--3754 reveal an X-ray spectrum quite close to that of a
blackbody of temperature $T= 61.2 \pm 1.0$ eV, and the data contain
evidence for neither spectral features nor pulsation.~\footnote{They
placed a 99$\%$ confidence upper limit of 2.7$\%$ on the unaccelerated
pulse fraction from 10$^{-4}$ to 100 Hz. Burwitz et
al.~\cite{Burwitz:2002vm} reported that a subsequent observation of RX
J1856.5--3754 with {\it XMM-Newton} allows the upper limit on the
periodic variation in the X-ray region to be reduced to 1.3$\%$ at
99$\%$ C.L. from 10$^{-3}$ to 50 Hz.} They argued that the derived
interstellar medium neutral hydrogen column density is
$N_H=(0.8$--$1.1)\times 10^{20}$ cm$^{-2}$, which, together with the
results of recent HST parallax analyses, yields an estimate of
111--170 pc for the distance $D$ to RX J1856.5--3754. Combining this
range of $D$ with the blackbody fit leads to a radiation radius of
$R_{\infty}=3.8$--8.2 km, which is smaller than typical neutron star
radii and thus suggests that the X-ray source may be a quark star.

In a subsequent work, Walter and Lattimer\cite{Walter:2002} showed
that the blackbody model of Drake et al.\ does not reproduce the
observed UV/optical spectrum.\cite{Pons:2001px} They succeeded in
fitting the two-temperature blackbody and heavy-element atmosphere
models developed in Ref.~\citen{Pons:2001px} to the observed spectrum,
ranging from X-ray to optical wavelengths. They found that the
radiation radius inferred from such a fitting lies between 12 km and
26 km, which is consistent with that of a neutron star. However, this
model is not effective for explaining the lack of spectral features.
Most recently, Braje and Romani\cite{Braje:2002} suggested that a
two-temperature blackbody model, which can reproduce both the X-ray
and optical-UV spectral data but seems to be inconsistent with the
fact that pulsation is not detected, is indeed compatible with the
interpretation that the object is a young normal pulsar, whose
nonthermal radio beam misses Earth's line of sight. However, this
model cannot answer why there are no features in the observed X-ray
spectrum. In the absence of a model that accounts for all the
observational facts, in this paper we employ the simple picture of
Drake et al.\cite{Drake:2002bj} based on the uniform temperature
blackbody fit to the X-ray spectrum.

{}From the radiation radius of RX J1856.5--3754 inferred from the
blackbody fit to the X-ray spectrum, the upper bound of the star's
mass can be derived.~\cite{prasanna:2002,Nakamura:2002,rosinska:2002}
This is because the radiation radius is larger than the true radius by
a factor arising from gravitational redshift effects. The analysis of
Drake et al. yields an upper bound of 0.5--$1.1 M_{\odot}$. In
previous studies,~\cite{prasanna:2002,rosinska:2002} the mass-radius
relations of pure quark stars were calculated from several kinds of
equations of state (EOS) for quark matter, and these results were
compared with the constraints on the mass and radius of RX
J1856.5--3754 derived from the inferred radiation radius. In this
framework, details of the relation between the mass of RX
J1856.5--3754 and the EOS remains to be clarified. Further systematic
investigations allowing for uncertainties in the EOS are needed; it is
expected that these uncertainties can be reduced using Monte-Carlo
calculations of lattice gauge theory at non-zero density, though such
calculations at present are beset by technical problems.

Taking into account the above points in this paper, we consider the
EOS of quark matter in the space of the parameters characterizing the
bag model and investigate the question of how massive a quark star
could be, given the inferred radiation radius. In \S2, we construct
quark stars using the bag-model EOS of quark matter. Section 3 is
devoted to derivation of the mass-radius relation and its application
to the soft X-ray source RX J1856.5--3754. Conclusions are given in
\S4. We use natural units in which $ \hbar = c = k_{\rm B} = 1$
throughout the paper.

\section{Quark star models}

To construct stars composed of zero-temperature $uds$ quark matter, we
start with the thermodynamic potentials, $\Omega_q$, for a homogeneous
gas of $q$ quarks ($q=u,d,s$) of rest mass $m_q$ to first order in
$\ac$, where $\ac \equiv g_{c}^{2}/4\pi$ is the fine structure
constant associated with the QCD coupling constant $g_{c}$. In
Refs.~\citen{Baym:1975va,Freedman:1976xs} and~\citen{Baluni:1977mk},
expressions for $\Omega_q$ are given as sums of the kinetic term and
the one-gluon-exchange term at the renormalization scale
$\Lambda=m_q$. We assume that $m_u=m_d=0$ and write $\Omega_q$ as
\begin{eqnarray}
    \label{eq:Omegau}
    \Omega_{u} &=& -
    \frac{\muu^{4}}{4\pi^{2}}\left(1-\frac{2\ac}{\pi}\right), \\ 
    \label{eq:Omegad}
    \Omega_{d} &=&
    - \frac{\mud^{4}}{4\pi^{2}}\left(1-\frac{2\ac}{\pi}\right), \\
    \label{eq:Omegas}
    \Omega_{s} &=&
    - \frac{m_s^{4}}{4\pi^{2}}\left\{
      x_s\eta_s^3-\frac32F(x_s) \right. \nonumber  \\ 
       && \left.  -\frac{2\ac}{\pi}\left[ 
          3F(x_s)\left(F(x_s)+2\ln{x_s} \right) -2\eta_s^{4}
          +6\left(\ln{\frac{\Lambda}{\mus}}\right)
          F(x_s) \right]
    \right\}, 
\end{eqnarray}
with
\begin{eqnarray}
  \label{eq:Fxs}
  F(x_s) = x_s\eta_s-\ln{(x_s+\eta_s)},
\end{eqnarray}
where $x_{s} = \mus/m_{s}$, $\eta_{s}=\sqrt{x_{s}^{2}-1}$, $\mu_{q}$
is the chemical potential of $q$ quarks, $\ac = \ac(\Lambda)$, and
$m_{s} = m_{s}(\Lambda)$. In this paper we choose $\Lambda =
\mu_s$.\footnote{ Farhi and Jaffe\cite{Farhi:1984qu} chose the
renormalization scale as $\Lambda=313$ MeV in describing self-bound
$uds$ quark matter in bulk; in such matter, the one-gluon-exchange
interaction energy is small compared with the kinetic energy. Here we
describe $uds$ quark matter forming a system of stellar size with a
much wider density range. The present choice ($\Lambda=\mu_s$) is
sufficient to keep the interaction energy relatively small for such a
density range. }

In a star, electrons are distributed in such a manner that each local
region is electrically neutral. (We ignore muons, which are more
massive than electrons.) Since generally the electrons are dense
enough to behave as an ultrarelativistic ideal gas, we can write the
electron thermodynamic potential $\Omega_e$ in the massless
noninteracting form,
\begin{eqnarray}
        \label{eq:Omegae}
    \Omega_{e} &=& - \frac{\mue^{4}}{12\pi^{2}},
\end{eqnarray}
where $\mue$ is the electron chemical potential.

In the bag model, we express the total energy density $\rho$ 
as\cite{Farhi:1984qu}
\begin{eqnarray}
    \label{eq:energy}
    \rho = \sum_{i=u,d,s,e}\left(\Omega_{i} +  \mu_{i}n_{i}\right) + B,
\end{eqnarray}
where $B$ is the bag constant, i.e., the excess energy density
effectively representing the nonperturbative color confining
interactions, and $n_i$ is the number density of $i$ particles, given
by
\begin{eqnarray}
    \label{eq:number}
    n_{i} = - \frac{\partial\Omega_{i}}{\partial\mu_{i}}.
\end{eqnarray}
Then, we obtain the pressure $p$ as
\begin{eqnarray}
    \label{eq:pressure}
    p &=& \mu_B n_{B} - \rho \nonumber \\ &=& -
    \sum_{i=u,d,s,e}\Omega_{i} - B,
\end{eqnarray}
where $\mu_B = \sum_{i=u,d,s,e}n_i \mu_i/n_B$ is the baryon chemical 
potential, and 
\begin{eqnarray}
    \label{eq:baryon_number}
    n_{B} \equiv \frac13\left(n_{u} +  n_{d} + n_{s} \right)
\end{eqnarray}
is the baryon density.

The values of the parameters $m_s$, $\ac$ and $B$ were obtained from
fits to light hadron spectra in
Refs.~\citen{DeGrand:cf,Carlson:er,Bartelski:1983cc} (see
Table~\ref{table:fitting}). However, the values yielded by such fits
do not correspond to bulk quark matter.~\cite{Farhi:1984qu} Allowing
for possible uncertainties, we consider values in the ranges $130\leq
B^{1/4}\leq 250$ MeV, $0\leq m_s \leq 300$ MeV, and $0\leq\ac\leq0.9$.
The ranges of $m_s$ and $\ac$ so chosen are not consistent in all
cases with the values in Table~\ref{table:fitting}, but they are
consistent with the Particle Data Group\cite{PDG} values renormalized
at an energy scale of interest here.

\begin{table}[htbp]
     \begin{center}
         \leavevmode
         \caption{
         Bag-model parameters fitted to hadron mass spectra.
         }
         \begin{tabular}{cccccc}
           \hline
             & $B^{1/4}$(MeV) & $m_{s}$ (MeV)& $\alpha_{c}$ &
             Reference& \\ \hline
             & 145
             &  279
             & 2.2
             & T.~DeGrand et al.\ (1975)~\cite{DeGrand:cf}
             &\\
             & 200--220
             & 288
             & 0.8--0.9
             & C.E.~Carlson et al.\ (1983)~\cite{Carlson:er}
             &\\
             & 149
             & 283
             & 2.0
             & J.~Bartelski et al.\ (1984)~\cite{Bartelski:1983cc}
             &\\
           \hline
         \end{tabular}
         \label{table:fitting}
     \end{center}
\end{table}

To obtain the equilibrium composition of quark matter in its ground
state at a given baryon density or chemical potential, the conditions
of equilibrium with respect to the weak interaction and of overall
charge neutrality must be imposed. These conditions are expressed by
\begin{eqnarray}
    \label{eq:weak-equiv1}
    \muu &=& \mud - \mue, \\
    \label{eq:weak-equiv2}
    \mud &=&  \mus, 
\end{eqnarray}
and 
\begin{eqnarray}
    \label{eq:charge_neutral}
    \frac23n_{u} - \frac13n_{d} - \frac13 n_{s} =  n_{e}.
\end{eqnarray}
{}From Eqs.~(\ref{eq:Omegau})--(\ref{eq:charge_neutral}), we can
obtain the energy density and the pressure as functions of $n_{B}$ (or
$\mu_B$).

\begin{figure}[htbp]
\centerline{\epsfxsize=0.65\textwidth\epsfbox{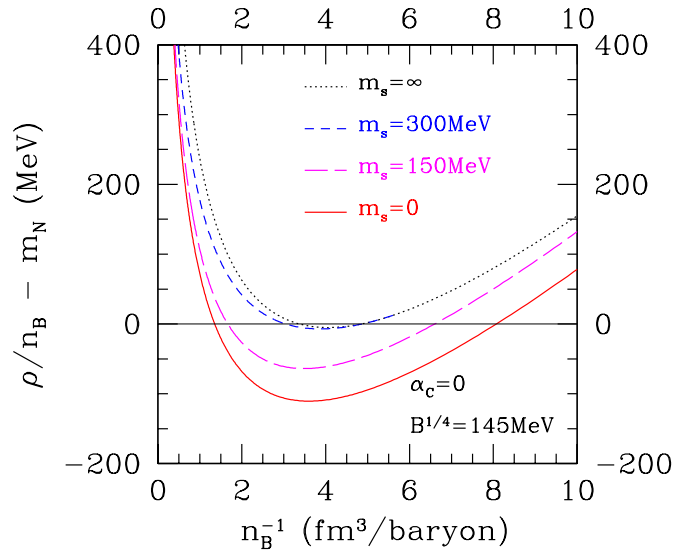}}
\caption{ Energy per baryon of electrically neutral quark matter in
$\beta$ equilibrium (minus the nucleon rest mass $m_N$) as a function
of $n_{B}^{-1}$, calculated for $\ac = 0$ and $B^{1/4} = 145$ MeV. The
dotted curve is the result for $ud$ quark matter ($m_{s} = \infty$).
The dashed (long dashed) curve is the result for $uds$ quark matter
with $m_s=300$ (150) MeV. The solid curve represents the result for a
massless three-flavor system ($m_{s} = 0$). }
\label{fig:EN_VNa0.0}
\end{figure}

\begin{figure}[htbp]
\centerline{\epsfxsize=0.65\textwidth\epsfbox{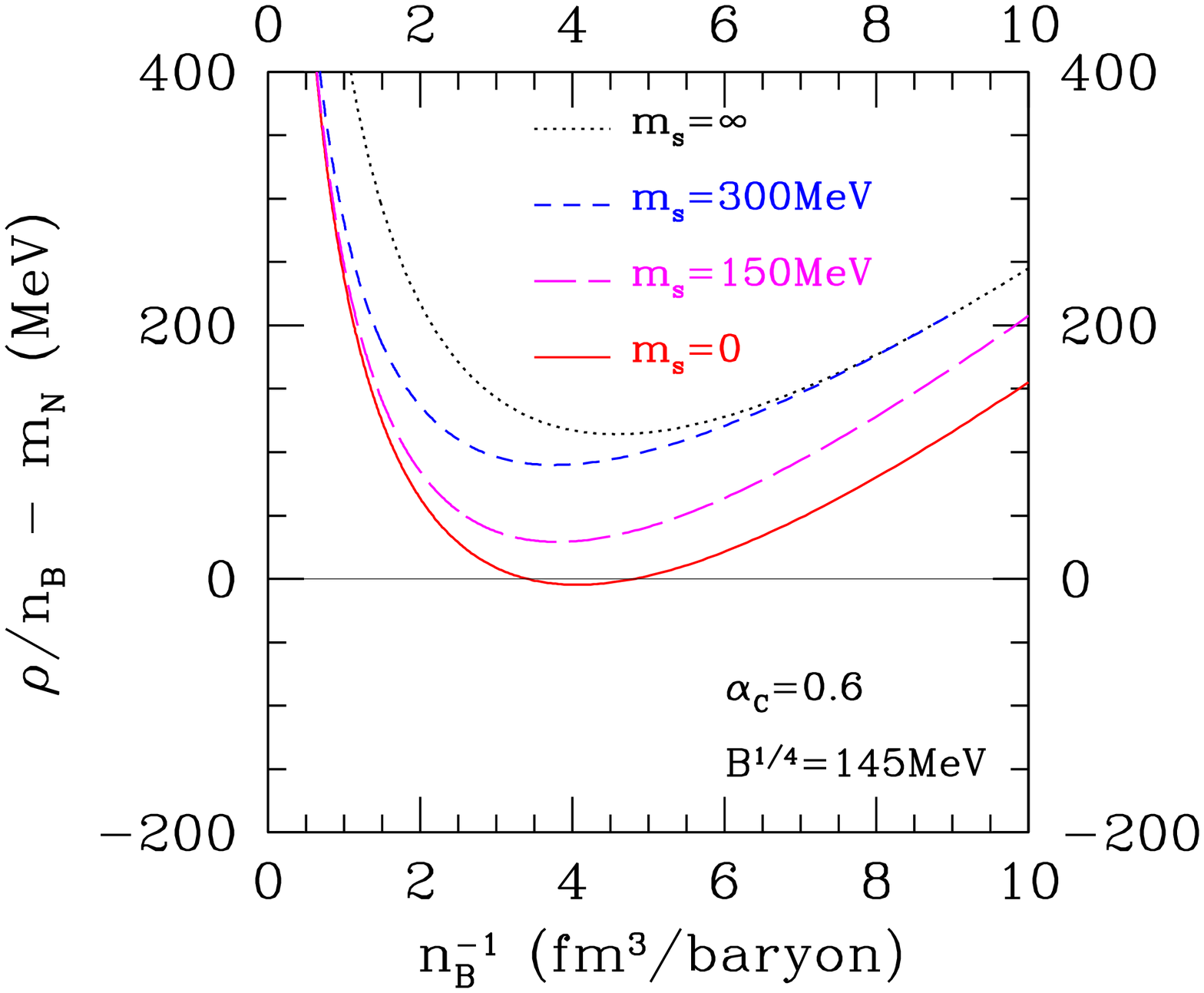}}
\caption{
Same as Fig.~\ref{fig:EN_VNa0.0}, but $\ac = 0.6$.
}
\label{fig:EN_VNa0.6}
\end{figure}

In Figs.~\ref{fig:EN_VNa0.0} and \ref{fig:EN_VNa0.6}, we plot the
energy per baryon ($\rho/n_{B}$) as a function of $n_{B}^{-1}$. We
note that $\rho/n_{B}$ becomes large as the strange quark mass $m_s$
increases, because $\mu_s$ is an increasing function of $m_s$ for
fixed $n_s$. We also find that $\rho/n_{B}$ increases as $\alpha_c$
increases. This is because the one-gluon-exchange interaction is
repulsive, due to the contribution from the exchange of transverse
(magnetic) gluons in a relativistic quark plasma.

\begin{figure}[htbp]
\centerline{\epsfxsize=0.65\textwidth\epsfbox{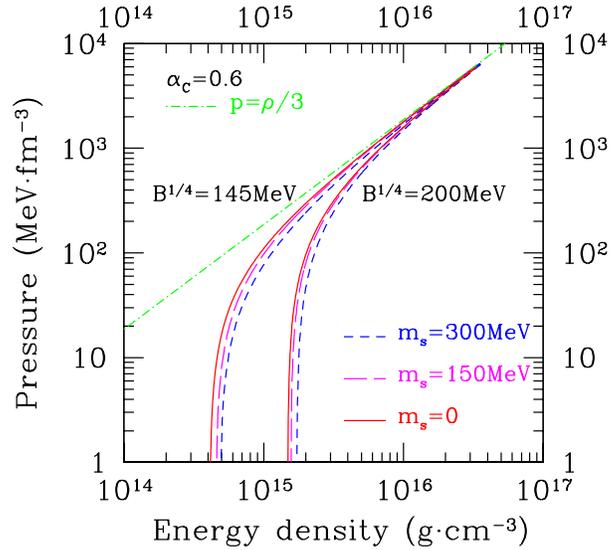}}
\caption{ Equation of state of $uds$ quark matter in the ground
state, calculated for $B^{1/4} = 145, 200$ MeV and $\ac = 0.6$. The
dashed, long dashed, and solid curves are the results for $m_s = 300,
150, 0$ MeV, respectively. The long dashed curve represents the EOS of
a massless ideal gas ($p = \rho/3$). }
\label{fig:rho_p}
\end{figure}

Figure \ref{fig:rho_p} depicts the equation of state of ground-state
quark matter, i.e., the pressure as a function of the energy density.
{}From this figure, we see that the EOS becomes softer as $B$
increases and hence as the color confining force increases. Such a
feature is explicitly expressed in Eq.~(\ref{eq:pressure}). We also
find that the EOS becomes softer as $s$ quarks become more massive.
This softening stems from the fact that a nonzero but small $m_s$ acts
to reduce the kinetic pressure of $s$ quarks, as can be seen from
Eq.~(\ref{eq:Omegas}). We remark in passing that as $\rho$ increases
the quark matter EOS becomes closer and closer to the massless
ideal-gas limit ($p = \rho/3$).

We proceed to calculate the structure of nonrotating quark stars as in
Refs.~\citen{Alcock:1986hz} and \citen{Haensel:qb}.  Such calculations can be 
performed by incorporating the EOS models obtained above into the general 
relativistic equation of hydrostatic equilibrium, i.e.,
the Tolman-Oppenheimer-Volkoff (TOV) equation,\cite{Oppenheimer:1939ne}
\begin{eqnarray}
    \label{eq:TOV}
    \frac{dp(r)}{dr} = - G\frac{\left[\rho(r) +
      p(r)\right]\left[M(r)+4\pi
      r^{3}p(r)\right]}
    {r^2\left[1-\displaystyle{\frac{2GM(r)}{r}}\right]},
\end{eqnarray}
where $G$ is the gravitational constant, $r$ is the radial coordinate
from the center of the star, and $M(r)$ is the gravitational mass of
the stellar portion inside a surface of radius $r$, which can be
obtained by integrating the equation for mass conservation,
\begin{eqnarray}
    \label{eq:mass_conv}
    \frac{dM(r)}{dr} = 4 \pi r^{2}\rho(r),
\end{eqnarray}
from $0$ to $r$. Then we can determine the radius, $R$, of the star
from the condition that the pressure becomes zero, i.e., $p(r=R) = 0$.
For various values of the energy density, $\rho_{0}$, at the center of
the star, we finally obtain the radius $R$ and the mass $M\equiv
M(R)$.

\begin{figure}[htbp]
\centerline{\epsfxsize=0.65\textwidth\epsfbox{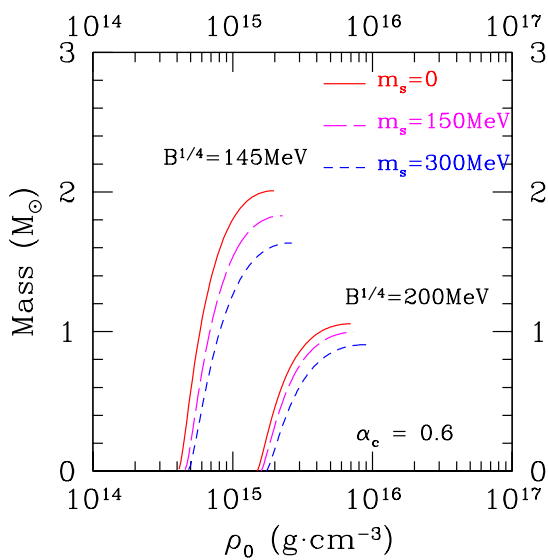}}
\caption{ Masses of pure quark stars with central energy density
$\rho_{0}$, calculated for $B^{{1/4}} = 145, 200$ MeV and $\ac = 0.6$.
The dashed, long dashed, and solid curves are the results for $m_s =
300, 150, 0$ MeV, respectively. }
\label{fig:rho0_M}
\end{figure}

\begin{figure}[htbp]
\centerline{\epsfxsize=0.65\textwidth\epsfbox{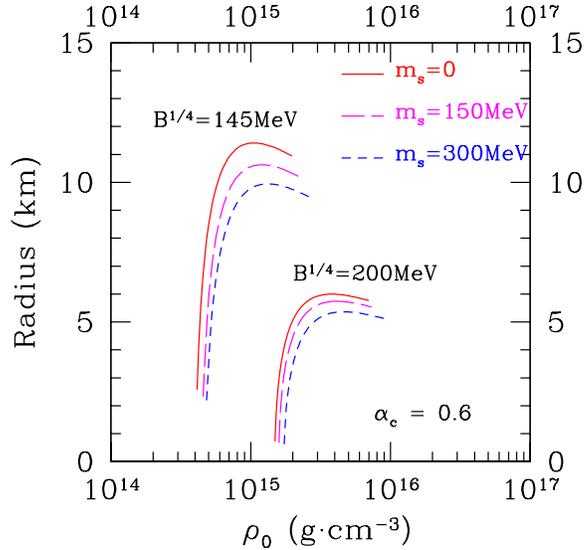}}
\caption{
Same as Fig.~\ref{fig:rho0_M}, except that the quantity plotted is
the radius rather than the mass.
}
\label{fig:rho0_R}
\end{figure}

In Fig.~\ref{fig:rho0_M} we plot the mass $M$ as a function of
$\rho_{0}$. In this figure, we include the $M$-$\rho_0$ relation up to
the maximally allowed value of the mass; at this point, the star is no
longer gravitationally stable.\cite{shapiro:1983} We also exhibit the
radius $R$ as a function of $\rho_{0}$ in Fig.~\ref{fig:rho0_R}, where
the right end of the $R$-$\rho_0$ relation corresponds to the maximum
mass for a quark star~\footnote{ For $m_s=0$, the mass, radius and
central energy density of a stable quark star of maximum mass are
$M=2.00M_{\odot}(B_0/B)^{1/2}$, $R=11.1$ km $(B_0/B)^{1/2}$ and
$\rho_0=1.91\times10^{15}$ g cm$^{-3}$ $(B/B_0)$.\cite{Witten:1984rs}
}. As can be seen in both Figs.~\ref{fig:rho0_M} and \ref{fig:rho0_R},
the radius and the mass become smaller for fixed $\rho_{0}$ as $B$
and/or $m_s$ increases, which leads to softening of the EOS (see Fig.\
\ref{fig:rho_p}). In other words, an increase in $B$ and/or $m_s$ for
a fixed radius or mass tends to increase $\rho_{0}$.

\section{Upper limit on the mass of the possible quark star}

\begin{figure}[htbp]
\centerline{\epsfxsize=0.65\textwidth\epsfbox{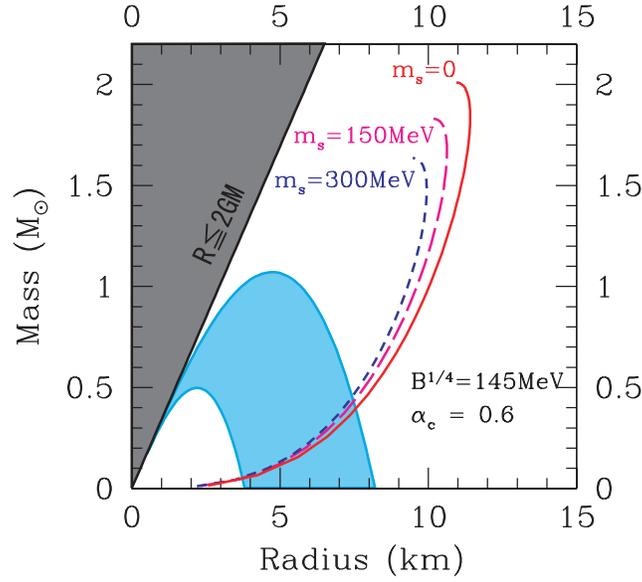}}
\caption{ Mass-radius relations of pure quark stars, calculated for
$\ac = 0.6$ and $B^{1/4} = 145$ MeV. The dashed, long dashed, and
solid curves are the results for $m_s = 300, 150, 0$ MeV,
respectively. The light shadowed region is allowed by the inferred
radiation radius, $R_{\infty} = 3.8$--8.2 km, of RX J1856.5--3754. The
dark shadowed region is excluded by the condition that $R$ is larger
than $2G M$. }
\label{fig:a0.6_B145}
\end{figure}

\begin{figure}[htbp]
\centerline{\epsfxsize=0.65\textwidth\epsfbox{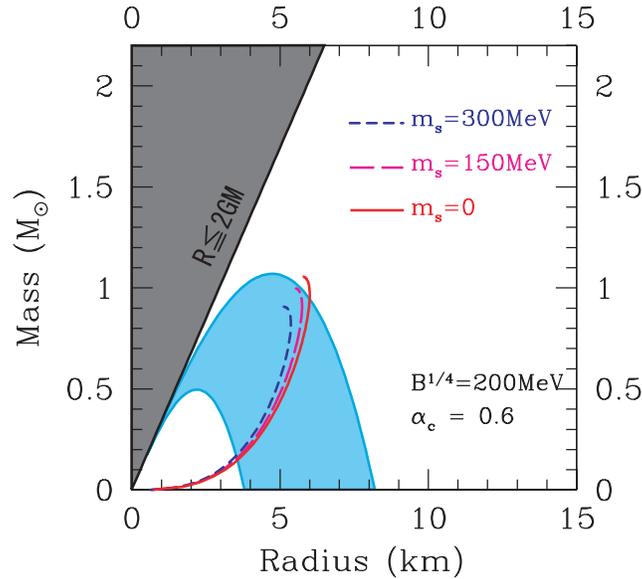}}
\caption{
Same as Fig.~\ref{fig:a0.6_B145}, but for $B^{1/4} = 200$ MeV.
}
\label{fig:a0.6_B200}
\end{figure}

In this section we derive the mass-radius relations of pure quark
stars for various bag-model parameters and compare these relations
with the inferred radiation radius of the soft X-ray source RX
J1856.5--3754. We can straightforwardly calculate the mass-radius
relations from the $M$-$\rho_0$ and $R$-$\rho_0$ relations obtained in
the previous section. The results for $B^{1/4} = 145, 200$ MeV, $m_s =
0, 150, 300$ MeV, and $\alpha_c=0.6$ are plotted in
Figs.~\ref{fig:a0.6_B145} and \ref{fig:a0.6_B200}. The $M$-$R$
relation ends at the maximum mass, where the gravitational instability
sets in, while the mass $M$, when small, behaves as $\sim R^3$, due to
the vacuum pressure $B$.

We now compare the obtained mass-radius relations of quark stars with
the inferred radiation radius of RX J1856.5--3754. According to the
analysis of the deep Chandra LETG+HRC-S observation of RX
J1856.5--3754 carried out by Drake et al.,\cite{Drake:2002bj} the
X-ray spectrum is quite close to that of a blackbody of temperature $T
= 61.2 \pm 1.0$ eV, and the X-ray luminosity is $L_X \simeq 6 \times
10^{31} (D/140~{\rm pc})^2$ erg s$^{-1}$.\cite{Drake:2002bj} Then, we
obtain the relation, $L_X = 4\pi R_{\infty}^2
\left(\pi^2/60T^4\right)$ between the luminosity and the temperature
with the radiation radius $R_{\infty}$ given by
\begin{eqnarray}
    \label{eq:radradi}
    R_{\infty} = \frac{R}{\sqrt{1 - \displaystyle{\frac{2 GM}{R}}}}.
\end{eqnarray}
The factor $1/\sqrt{1-2GM/R}$ represents the redshift effect in a
strong gravitational field. Drake et al.\cite{Drake:2002bj} inferred
$R_{\infty} =3.8$--8.2 km from the distance to RX J1856.5--3754, $D$=
111--170 pc.

Given such an inferred radiation radius $R_{\infty}$, we can obtain
information about the radius $R$ and the mass $M$ of the star from
Eq.~(\ref{eq:radradi}). In Figs.~\ref{fig:a0.6_B145} and
\ref{fig:a0.6_B200}, the light shadowed region denotes the region of
$R$ and $M$ allowed by the radiation radius $R_{\infty} = 3.8$--8.2
km. The dark shadowed region in these figures is excluded by the
condition that the radius $R$ of a radiation emitter be larger than a
black hole surface of radius $2 GM$. {}From these figures, we can see
that the mass upper limit, $M_{\rm up}$, allowed by $R_{\infty}$
increases from $\sim 0.4 M_{\odot}$ to $\sim 1 M_{\odot}$ when
$B^{1/4}$ increases from 145 MeV to 200 MeV, while being almost
independent of $m_s$. We remark that the mass lower limit allowed by
$R_{\infty}$ is of order $0.1 M_{\odot}$.

\begin{figure}[htbp]
\centerline{\epsfxsize=0.9\textwidth\epsfbox{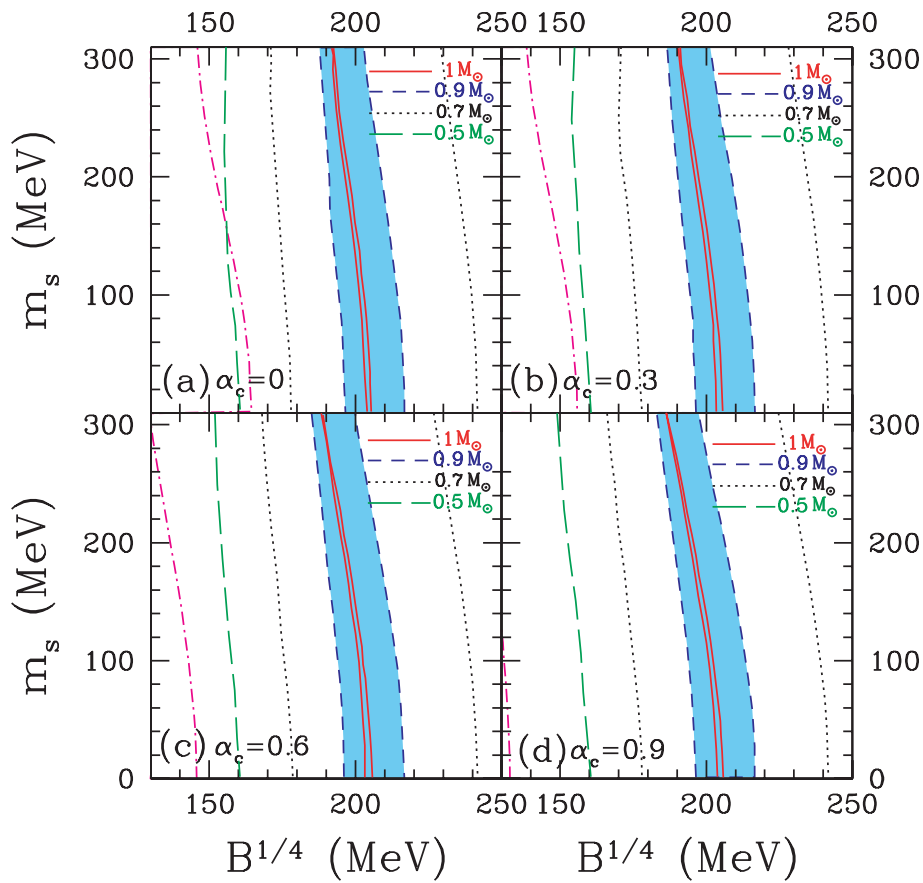}}
\caption{ Contours of the mass upper limit, $M_{\rm up}$, allowed by
the inferred radiation radius of the soft X-ray source RX
J1856.5--3754 in $B^{1/4}$--$m_s$ plane, calculated for (a)
$\alpha_c$= 0, (b) 0.3, (c) 0.6, and (d) 0.9. In each panel the
contours for $M_{\rm up}=1 M_{\odot}$ (solid curves), $M_{\rm up}=0.9
M_{\odot}$ (dashed curves), $M_{\rm up}=0.7 M_{\odot}$ (dotted
curves), and $M_{\rm up}=0.5 M_{\odot}$ (long dashed curves) are
included. For reference, we denote by the dot-dashed curves the
boundary below which strange matter is absolutely stable at zero
pressure (for details, see Fig.~1 in Farhi and
Jaffe~\cite{Farhi:1984qu}). }
\label{fig:panel}
\end{figure}

In Fig.~\ref{fig:panel} we plot the contours of the mass upper limit
$M_{\rm up}$ in the space of the parameters $B^{1/4}$ and $m_s$,
calculated for $\alpha_c= 0, 0.3, 0.6, 0.9$. {}From this figure, we
find that the contours are peaked around $B^{1/4}\sim200$ MeV. The
left side of this peak results from the boundary of the allowed
$M$-$R$ region by the inferred radiation radius of RX J1856.5--3754,
while the right side is due to the maximum mass of gravitationally
stable quark stars. We remark that for the smallest values of $B$, in
addition to strange matter, even $ud$ quark matter at zero pressure
can be more stable than normal nuclei, as shown by Farhi and
Jaffe.\cite{Farhi:1984qu} We also note that the absolute stability of
strange matter at zero pressure, which should allow for the existence
of strange stars to exist, limits the bag-parameter region to the left
side of the dot-dashed curves. It might be possible to relax this
limitation if we consider that the bag constant depends on the
pressure, as discussed in the Appendix.

\section{Conclusions}

In this paper we have calculated the mass-radius relations of pure
quark stars for various sets of bag-model parameter values. Assuming
that RX J1856.5--3754 is a pure quark star, we have derived an upper
limit on its mass, which is of order 0.5--$1 M_{\odot}$. In such a
star of $M\simeq 1 M_{\odot}$, the central energy density would be
$\sim 10^{16}$ g cm$^{-3}$. The present systematic analysis
essentially includes the cases treated in previous analyses utilizing
the EOS models based on perturbation theory\cite{Fraga,Andersen} and
finite-temperature lattice data.\cite{Peshier} Because this paper
treats pure quark stars alone, however, many questions remain. We now
discuss some of these.

Unless $uds$ quark matter forms a more stable self-bound system than
$^{56}$Fe, a pure quark star would likely be unstable with respect to
hadronization. The resultant star, no matter whether it be a hybrid
star or a member of the third family mentioned in the Introduction,
should have a larger radius for fixed central energy density with a
hadronic shell surrounding a quark matter core. The extent to which
the radius would increase depends strongly on the model of dense
hadronic and quark matter used. Due to this increase, it is possible
that the situation in which the bag-model parameter sets do not yield
the absolute stability of $uds$ quark matter would be in contradiction
with the inferred radiation radius of RX J1856.5--3754, except for the
case of possible parameter sets including large $B$ for which the
mass-radius relation of the third family enters the region allowed by
the inferred radiation radius.

The possibility for the existence of a quark star of the third family
depends on the condition that the velocity of sound be drastically
enhanced at a deconfinement transition.~\cite{Gerlach} It may be
possible for this condition to be satisfied, as shown in
Refs.~\citen{Glendenning} and \citen{Schertler} in the case in which
the deconfinement transition is accompanied by a quark-hadron mixed
phase including $uds$ quark droplets embedded in a gas of electrons,
in a sea of hadronic matter. This phase may be realizable because the
presence of $s$ and $d$ quarks acts to reduce the electron Fermi
energy and to enhance the proton fraction in hadronic matter. However,
there are several reasons why it might be difficult for such a mixed
phase to appear. First, the quark-hadron interfacial energy, which is
poorly known, might be large enough to make the mixed phase
energetically unfavorable.\cite{Heiselberg} Second, even for fairly
small values of the interfacial energy, electron screening of quark
droplets could act to destabilize the mixed phase.\cite{Yasuhira}
Finally, even if the mixed phase is energetically favorable, it is
uncertain how the mixed phase would nucleate in a neutron
star.~\cite{Iida}

If $uds$ quark matter is absolutely stable, strange stars could exist,
with a possible crust of nuclear matter. Such a crust would likely be
composed of a Coulomb lattice of nuclei immersed in a roughly uniform
gas of electrons, but not in a sea of neutrons.~\cite{Alcock:1986hz}
This nuclear matter including free neutrons would be surrounded by
absolutely stable $uds$ quark matter; the electric charge of the quark
region, which is generally positive, would prevent the normal nuclei
from contacting the quark matter. The effects of the crust on the
strange star structure were investigated by Zdunik,~\cite{Zdunik} He
showed that the maximal presence of the crust would qualitatively
change the mass-radius relation only for $M\siml 0.1 M_{\odot}$.
Determining to what extent the crust would prevail in a strange star
requires information about the formation and evolution of the star. We
remark that even if RX J1856.5--3754 is a strange star, the surface of
the star should be hadronic. This is because the observed spectrum is
of blackbody type, while at the observed X-ray luminosity, the
spectrum emitted from a bare quark matter surface would differ
drastically from the blackbody spectrum.~\cite{Page}

Information about the mass of RX J1856.5--3754 is useful to
distinguish between the possible forms of compact stars. In order to
derive a constraint on the mass from the observed X-ray luminosity
$L_X$, it is instructive to note the possible origins of $L_X$.
Initial cooling and accretion of interstellar material are generally
considered to be responsible for $L_X$.\cite{Walter:1996} We may thus
assume that the gravitational energy release of the accreting material
is smaller than $L_X$, i.e., $L_X \simg GM\dot{M}/R$, with the Bondi
accretion rate $\dot{M}$ onto a star moving supersonically in an
ambient medium given by~\cite{Bondi:1944}
\begin{eqnarray}
    \label{eq:Bondi_acc}
    \dot{M} = 4\pi \lambda
    \left(\frac{G M}{v^2}\right)^2\rho_H v.
\end{eqnarray}
Here, $\lambda$ is an $\order(1)$ constant related to the EOS of the
accreting matter, $v$ is the velocity of the star, and $\rho_H$ is the
hydrogen mass density ($= m_N n_H$ with nucleon mass $m_N$ and
hydrogen number density $n_H$). The velocity $v$ has been determined
as $v \simeq 200 (D/140$ pc) km s$^{-1}$ from observations of the
proper motion of the optical counterpart of RX
J1856.5--3754.\cite{Walter:2002} In order to estimate $n_H$, we follow
the argument of Ref.~\citen{Nakamura:2002}. The hydrogen number
density is roughly given by $n_H = N_H/R_H$, where $N_H$ is the column
density to the star derived from the blackbody fit to the X-ray
spectrum as $N_H=(0.8$--$1.1)\times 10^{20}$
cm$^{-2}$,\cite{Drake:2002bj} and $R_H$ is the size of the local high
density region. $R_H$ could be on the order of 100 AU or even less, as
is known from observations of the interstellar medium using the 21 cm
H{\footnotesize I} line.\cite{Frail:1994} {}From
Eq.~(\ref{eq:Bondi_acc}), we can thus estimate the mass of the star as
\begin{eqnarray}
    \label{eq:mass_est}
    M &\siml& 0.4 M_{\odot} \left(\frac{v}{200~{\rm km}}\right)
    \left(\frac{D}{140~\pc}\right)^{5/3} \left(\frac{N_H}{10^{20}~
      \cm^{-2}}\right)^{-1/3} \left(\frac{R_H}{100~{\rm AU}}\right)^{1/3}
   \nonumber \\ & &
     \times\left(\frac{R}{5~{\rm km}}\right)^{1/3}.
\end{eqnarray}
Uncertainties in Eq.\ (\ref{eq:mass_est}) come mainly from the
velocity $v$ and the local high density scale $R_H$. The actual values
of $v$ and $R_H$ depend on the poorly known radial velocity and
density profile in the R CrA molecular cloud believed to contain RX
J1856.5--3754. Therefore, $M$ may in fact be much larger than
$0.4M_{\odot}$. However, it is interesting to note that the mass range
estimated from the X-ray luminosity $L_X$, which is almost independent
of the spectral analysis leading to the allowed $M$-$R$ region
depicted in Figs.\ \ref{fig:a0.6_B145} and \ref{fig:a0.6_B200}, may
play a role in narrowing the allowed $M$-$R$ region and hence in
clarifying the composition of this compact star.

Elucidating the formation of quark stars is another important problem.
If quark stars of masses smaller than $\sim 1 M_{\odot}$ are converted
from neutron stars with canonical masses of order $1.4 M_{\odot}$,
material in a neutron star should be partly ejected out of the star
and partly changed into quark matter. Thus, possible scenarios for the
formation of low-mass quark stars (some of which are discussed in
Refs.~\citen{Ouyed} and~\citen{Nakamura:2002}) depend inevitably on
the initial and final configuration of the stars.

\section*{Acknowledgments}
K.~K. wishes to thank Takashi Nakamura, T. Shigeyama, Shoichi Yamada,
S. Nagataki, and G. Watanabe for useful discussions and comments. This
work was supported in part by Grants-in-Aid for Scientific Research
provided by the Ministry of Education, Culture, Sports, Science and
Technology of Japan through Research Grant Nos.\ S14102004 and
14079202, and in part by RIKEN Special Postdoctoral Researchers Grant
No. 011-52040.

\appendix
\section{Effects of the Pressure Dependent Bag Constant}

In this appendix, we modify the bag model adopted in the main text by
allowing the bag constant to depend on the pressure. In the model thus
modified, it is possible for the bag constant inside the star to
become effectively larger than that at the surface. Then, the absolute
stability condition of strange matter at zero pressure, which is
characterized by the dot-dashed curves in Fig.~\ref{fig:panel}, can be
changed significantly. In the main text, we studied the mass-radius
relations of pure quark stars, with $B$ fixed throughout the
interiors, and extrapolated our calculations beyond the absolutely
stable region. As we see below, however, the stable region can be
enlarged by the pressure dependence of the bag constant.

As discussed in Refs.~\citen{Peshier} and \citen{Sinha}, we can
generally parameterize the EOS of strange matter in the massless limit
($m_s \to 0$) as
\begin{eqnarray}
    \label{mod_eos}
    p = \frac1{A}\left(\rho - 4 B \right),
\end{eqnarray}
where $A$ is a parameter, whose value is three in the standard bag
model. Then, we can transform Eq.\ (\ref{mod_eos}) into
\begin{eqnarray}
    \label{mod_eos2}
     p = \frac{1}{3} (\rho-4\beff),
\end{eqnarray}
where $\beff$ is the effective bag constant, which depends on the
pressure $p$ (or the energy density $\rho$) as
\begin{eqnarray}
    \label{beff}
    \beff &\equiv& \frac14\left(A - 3\right)p + B, \nonumber \\ &=&
    \frac14 \left(1-\frac{3}{A}\right) \rho +
    \left(\frac{3}{A}\right)B.
\end{eqnarray}
$\beff$ reduces to $B$ in the limit $p\to0$. {}From Eq.~(\ref{beff})
we find that if $A$ is larger than three, $\beff$ becomes larger than
$B$ inside the star. In the model used in Ref.~\citen{Peshier}, $A$
takes a typical value of 3.8, while $A \approx 2.2$ in Ref.\
\citen{Sinha}. Here we consider the range of values $A=$ 2--6. We
remark that there should be an upper bound on $A$; at non-zero
pressures, if $A$ is far greater than unity, the effective bag
constant is increased to such an extent that hadronic matter can be
energetically favored even at extremely high pressures.

\begin{figure}[htbp]
\centerline{\epsfxsize=0.6\textwidth\epsfbox{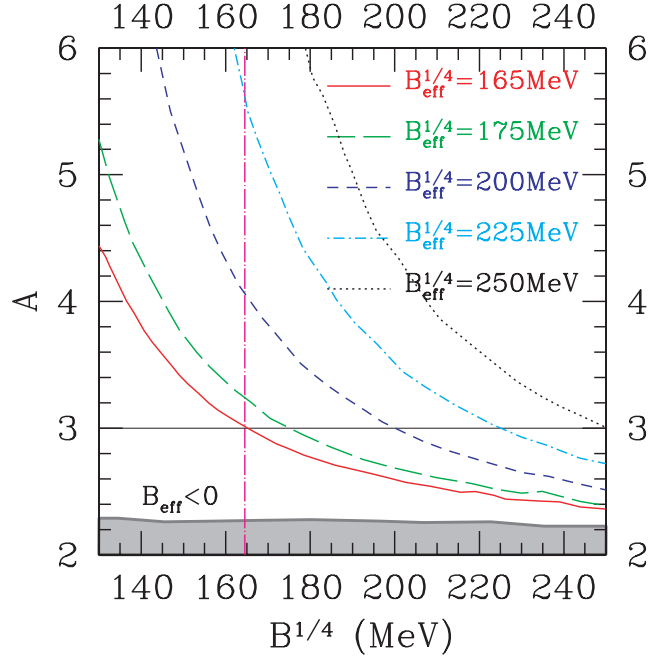}}
\caption{ Contours of the effective bag constant $\beff^{1/4}$ at the
central pressure of a stable quark star of maximum mass in the
$B^{1/4}$--$A$ plane, calculated for $\ac=0$. The vertical dot-dashed
curve is the boundary of the region in which strange matter is
absolutely stable at zero pressure ($B^{1/4} \le$ 165 MeV). The
shadowed region corresponds to $\beff \le 0$.}
\label{fig:beff}
\end{figure}

We now consider calculations of mass-radius relations of pure quark
stars from the EOS (\ref{mod_eos}). By repeating the calculations
performed in \S2 for $\beff$ rather than $B$, we can compute the
central pressure of a stable quark star of maximum mass for various
values of $A$ and $B$. In Fig.~\ref{fig:beff}, we plot the contours of
the effective bag constant $\beff^{1/4}$ at this central pressure in
the $B^{1/4}$--$A$ plane. We denote by the dot-dashed curve the
boundary of the region in which strange matter at the surface ($p=0$)
is absolutely stable ($B^{1/4} \le$ 165 MeV). {}From this figure, we
see that at a critical value of $B^{1/4}$ = 165 MeV, $\beff$ can be as
large as 200 MeV inside the star for $A \simg 4$. Since such large
values of $\beff$ effectively soften the EOS of strange matter, the
bag model with a pressure dependent bag constant allows for the
possibility of strange stars that are more compact than those
predicted by the bag model, as plotted in Fig.~\ref{fig:a0.6_B145},
for fixed $B$.

\begin{figure}[htbp]
\centerline{\epsfxsize=0.65\textwidth\epsfbox{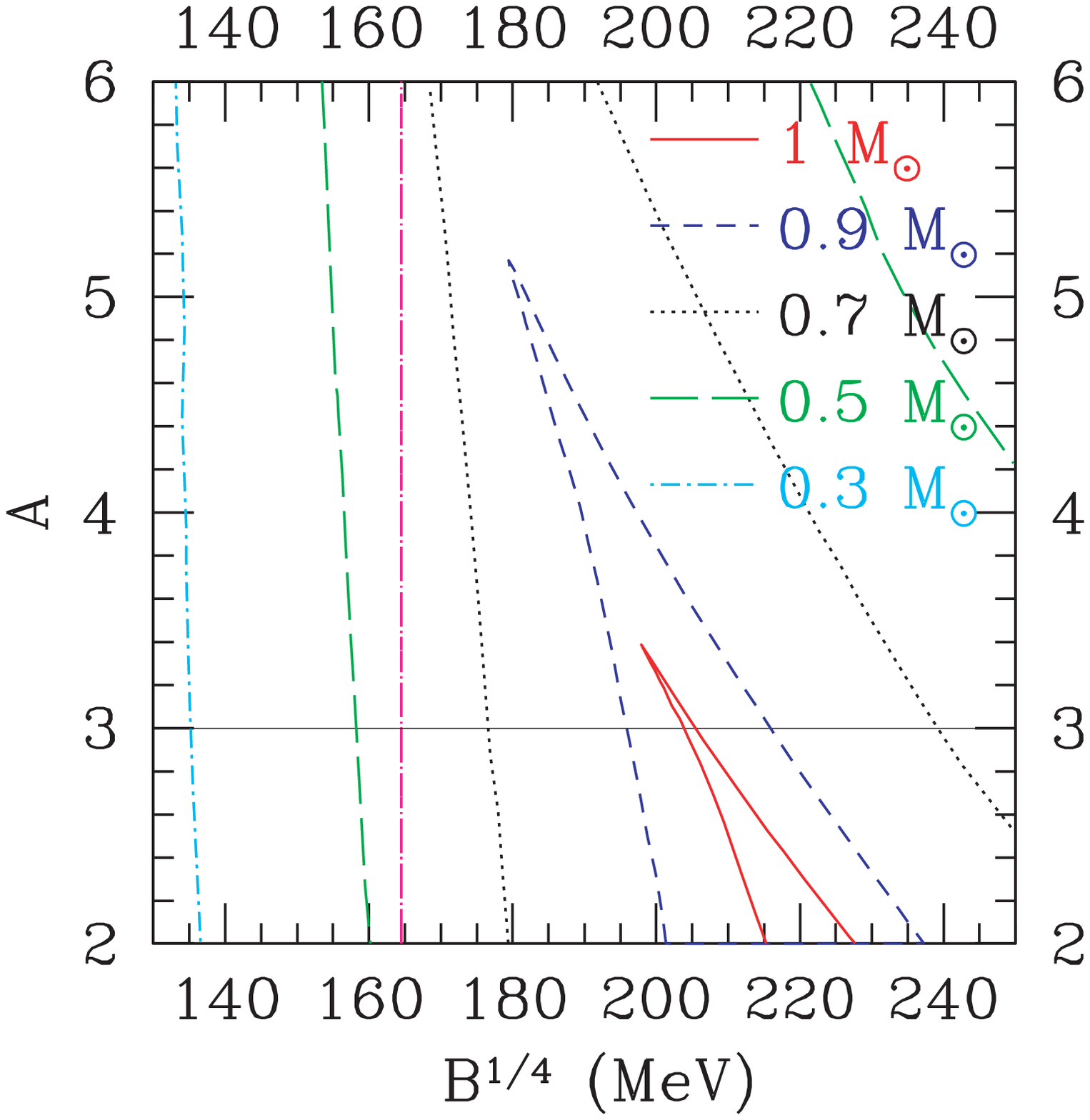}}
\caption{ Contours of the upper limit of the mass, $M_{\rm up}$,
allowed by the inferred radiation radius of the soft X-ray source RX
J1856.5--3754 in the $B^{1/4}$--$A$ plane, calculated for $\ac=0$. The
contours of $M_{\rm up}=1 M_{\odot}$ (solid curves), $M_{\rm up}=0.9
M_{\odot}$ (dashed curves), $M_{\rm up}=0.7 M_{\odot}$ (dotted
curves), $M_{\rm up}=0.5 M_{\odot}$ (long dashed curves) and $M_{\rm
up}=0.3 M_{\odot}$ (dot-short-dashed curves) are included. The
vertical dot-dashed curve is the boundary of the region in which
strange matter is absolutely stable at zero pressure ($B^{1/4} \le$
165 MeV). }
\label{fig:b4_A}
\end{figure}

In order to see how the resultant softening of the strange matter EOS
affects the upper limit of the mass, $M_{\rm up}$, allowed by the
inferred radiation radius of the soft X-ray source RX J1856.5--3754,
we plot the contours of $M_{\rm up}$ in the $B^{1/4}$--$A$ plane (see
Fig.~\ref{fig:b4_A}). When $A>3$, quark stars which are more massive
than those allowed by the bag model prediction ($A=3$) exist in the
region of the absolute stability of strange matter at zero pressure.
Because the inferred radiation radius of RX J1856.5--3754 is very
small, however, the change is small for $A\sim4$.

\end{document}